# Nano-corrugation induced forces between electrically neutral plasmonic objects

*Kun Ding,[†] Han Hu,[‡] T. C. Leung,[‡] and C. T. Chan[\*,†]*

[†] Department of Physics and Institute for Advanced Study, The Hong Kong University of Science and Technology, Clear Water Bay, Hong Kong

[‡] Department of Physics, National Chung Cheng University, Chia-Yi 62101, Taiwan

**ABSTRACT:** Recent advances in nanotechnology have created tremendous excitement across different disciplines but in order to fully control and manipulate nano-scale objects, we must understand the forces at work at the nano-scale, which can be very different from those that dominate the macro-scale. We show that there is a new kind of curvature-induced force that acts between nano-corrugated electrically neutral plasmonic surfaces. Absent in flat surfaces, such a force owes its existence entirely to geometric curvature, and originates from the kinetic energy associated with the electron density which tends to make the profile of the electron density smoother than that of the ionic background and hence induces curvature-induced local charges. Such a force cannot be found using standard classical electromagnetic approaches, and we use a self-consistent hydrodynamics model as well as first principles density functional calculations to explore the character of such forces. These two methods give qualitative similar results. We found that the force can be attractive or repulsive, depending on the details of the nano-corrugation, and its magnitude is comparable to light induced forces acting on plasmonic nano-objects.

Nano-machines have long been the theme of science fictions, and visionary scientists have predicted that nano-machines can perform tasks such as sensing, assembly, fabrication and even medical procedures in the not-too-distant future. In order to accomplish these tasks, it is essential to understand the forces at work at the microscopic level at the nano-scale. Nano-technology has advanced to a stage where nano-particles with designed morphologies and objects with precise nano-scale features can be fabricated by various top-down or bottom-up techniques. However, manipulating nano-scale objects[1] seems more complicated than making them. Nano-scale objects interact with each other in ways that are different from their macro counterparts and forces that are usually negligible for the macroscopic entities can become important at the nano-scale. For example, light induced forces are small compared to gravity for meter-scale objects, but at the micron and nanometer scale, optical forces can easily levitate or manipulate particles and have seen important applications in the form of optical tweezers.[2-8] For plasmonic systems, which we will consider in this paper, light induced forces can be very strong if the external driving field is tuned to excite plasmonic resonances, so that the field is strongly enhanced and confined near the nano-objects.[9-13] Charging nano-objects is another versatile and powerful way to control the interaction between nanostructured materials. Electrostatic forces[14] are most useful in colloidal environments and depending on experimental conditions, the range and the sign of the forces can be controlled.

But even in the absence of an external driving field or charging/gating, electrically neutral objects can naturally interact with each other through van der Waals and Casimir forces,[15-17] and these long range attractive forces can become significant when objects are close enough to each other. From the point of view of Lifshitz-type formalism,[18,19] these quantum fluctuation forces share the same origin, with Casimir forces being characterized by $\hbar c$, which highlight its quantum and electromagnetic origin. We note in passing that Casimir forces are usually perceived as an attractive force, but exceptions have been discovered for special configurations involving liquids,[20,21] artificial materials with unusual effective material dispersions[22-27] or surfaces with micro-protrusions.[28,29] In the nanometer scale we are interested in, retardation effects can be ignored and as such, quantum fluctuation forces can be labeled as van der Waals forces, which is attractive in nature.

In this paper, we show that there is a new kind of electrostatic force that exists between electrically neutral nano-corrugated plasmonic surfaces. This force is meaningful only at the nano-scale. The origin of this force can be traced to the kinetic energy of the electrons in the plasmonic system, and hence it can be considered quantum mechanical in nature. Simply put, the electronic wavefunctions tend to be smooth in order to minimize the positive definite kinetic energy of the electron gas and hence the profile of electron distribution tends to be smoother than the profile of ionic background in an object with protrusions or corrugations. On corrugated surfaces, the differences in the electronic and ionic profiles produce local charges, with a convex curvature favoring a deficit while a concave curvature favoring an excess of electrons. These local charges in turn can produce electric fields that can interact with other objects. While the origin is quantum mechanical, it is distinctly different from van der Waals forces which can be viewed as a second-order perturbation effect arising from fluctuations. The corrugation-induced force is a ground state effect due to the ground state electronic charge distribution that minimizes the energy of the electronic system, and has nothing to do with fluctuation. We will show that the corrugation-induced force can be attractive or repulsive, depending on geometric details. Such a force cannot be calculated using the classical electrodynamics approach, which will give a zero force between neutral plasmonic objects unless there is an external driving field. The method of calculation must include an energy functional for the electrons that contains a kinetic energy term. In this paper, we use two methods to calculate this force, namely a self-consistent hydrodynamic model (SC-HDM)[30-34] that incorporates the energy functional for the electrons and a standard local density functional approach (LDA) that naturally incorporates the kinetic energy of the electron wave functions. As a quantum-corrected classical model, SC-HDM is probably the minimal model that can describe the physics. It is applicable only to generic plasmonic systems, but it already captures the essence of the physics.[30-34] LDA is more accurate, and can be applied to just about any metallic system, but at a considerably higher computation cost. These two methods give qualitatively similar results that we shall describe in the following.

**Results and Discussion**

We employ the self-consistent hydrodynamics model (SC-HDM)[30-34] to study the forces between the charge neutral plasmonic slabs with nano-scale corrugations, as shown in Figure 1. From the viewpoint of classical electrodynamics, the force between neutral objects must be zero as there is no external driving field. We show that there is a corrugation induced electrostatic force between two electrically neutral objects. This geometry induced force originates from the interactions between the curvature-induced electrostatic dipoles (CEDs) near the corrugated surfaces, and its magnitude is comparable to the optical forces induced by an active external light source of fairly strong intensity (see section S1 in the Supporting Information for details). What makes this type of force interesting is that the force could be attractive or repulsive, depending on the details of the corrugation geometry.

**Corrugated Plasmonic Surfaces.** To understand the origin of the CEDs near the corrugated surfaces, we first consider a single metallic slab bounded by one flat interface and one corrugated interface, as shown in Figure 1a. The SC-HDM is employed to solve the ground state electron density $n_0$ and the related electrostatic potential $\phi_0$ generated by the electrons and nuclei[35] (see Methods for details). We are interested in understanding a generic geometric effect, and to that end, we employ a generic jellium model. To model a plasmonic slab using jellium, we need to define the boundaries of this positive charged jellium background, which is shown by the yellow regions in Figure 1a. We choose the analytical form of the jellium boundary for the corrugated surfaces as $y_{u,b} = \pm \frac{d}{2} + \frac{A_{u,b}}{2}\cos\left(\frac{\pi x}{r_a}\right)$. The subscript "u" ("b") stands for the upper (bottom) interface of the slab. Since the plasmonic slabs are periodic, we choose in the numerical calculations a unit cell that ranges from $x = -0.5 r_a$ to $x = +0.5 r_a$, where $r_a$ is the lattice constant. We apply boundary conditions in the x-direction, namely $n_0|_{x=-0.5 r_a} = n_0|_{x=+0.5 r_a}$ and $\phi_0|_{x=-0.5 r_a} = \phi_0|_{x=+0.5 r_a}$, while the boundary conditions in the y-direction is Dirichlet ($n_0|_{y=\pm\infty} = 0$). Since SC-HDM is good for simple metals,[30-34] we employ parameters that correspond to sodium. The ion density $n_{\text{ion}}$ of sodium is defined as $n_{\text{ion}} = \frac{3}{4\pi(r_s a_H)^3}$, where $a_H = 0.529\text{Å}$ is Bohr radius and dimensionless quantity $r_s = 4$.

**Charge Distribution from SC-HDM.** To facilitate our discussion, we define a normalized charge distribution as $\Delta n = (n_0 - n_j)/n_{ion}$, where $n_j$ is the jellium charge distribution ($n_j = n_{ion}$ inside the metal domain, and $n_j = 0$ for the vacuum domain). The normalized charge distribution for the cosine corrugated surface corresponding to the upper left panel of Figure 1a near the surface region is shown in Figure 2a. The parameters in Figure 2a are $r_a = 20 \text{Å}$, $d = 50 \text{Å}$, $A_u = 12 \text{Å}$, and $A_b = 0 \text{Å}$. We see that electrostatic dipoles exist near the surface, due to the well-known charge "spill-out" effect.[36] In order to distinguish this well-known effect with we are going to discuss below, we call this intrinsic surface dipole the interfacial electrostatic dipole (IED). While not obvious to the human eye, the IED as shown in Figure 2a is actually not homogenous on a corrugated surface. To see the difference of the IEDs between the convex and concave directions, we plot in Figure 2b the $\Delta n$ as a function of the y-coordinate (surface normal direction) at $x/r_a = 0.0$ and $x/r_a = 0.5$ by blue and red lines, respectively. Here, the zero of the y-coordinate is the jellium boundary. It is clear that the IEDs in these two directions are quite different. The electrons tend to concentrate inside the metal in the convex direction (blue line in Figure 2b), while the electrons in the concave direction tend to spill out from the jellium boundary (red line in Figure 2b). The physical origin of this difference comes from the kinetic energy term in the electron energy density functional. The kinetic energy term always tends to smooth out the electron charge distribution profile,[36] making it "less corrugated" than the ionic background charge distribution, hence rendering a concave valley having excess negative charge while a convex protrusion will have a deficit of negative charge. We note that this effect does not exist if we consider the system using classical electromagnetics, but it will show up in a quantum model that contains a kinetic energy term in the electron density functional.

To see quantitatively the difference between different directions, we define the surface charge density $\sigma(x) = \int \left[ e(n_j - n_0) \right] dy$ ($e > 0$) to characterize these IEDs. If the positive and negative charges inside these surface electrostatic dipoles exactly compensate each other, then $\sigma = 0$, indicating that the strength of the monopole component is zero and the leading order is the electric dipoles. If $\sigma > 0$ ($< 0$), the net surface charge of the local point is positive (negative), implying that the local monopole component is nonzero. In Figure 2d, we plot $\sigma$ as a function of x-coordinate for different values of $A_u$. When the surface is flat ($A_u = 0 \text{Å}$), $\sigma$ equals to zero as it must be (gray lines in Figure 2d), indicating that no net charge can accumulate near the flat surface. In other words, no monopole components exist for the flat interface between metal and vacuum. However, when we introduce corrugations to the flat surface ($A_u$ non-zero), some positive net charges can accumulate near the convex region, and some negative charges accumulate near the concave region, as shown by blue, red, green and magenta lines in Figure 2d. The non-zero surface charge $\sigma$ along the x-direction indicates that the local monopole component exists at each point. Furthermore, these local nonzero monopoles (surface charges) could form a new electrostatic dipole pointing from the concave region to the convex region, as shown by the inset in Figure 2d. This electrostatic dipole is different from the ones shown in Figure 2a. As mentioned earlier, the IEDs in Figure 2a exist in any interfaces, but the electrostatic dipoles shown in Figure 2d are caused by the geometric curvatures and exist in any nano-corrugated plasmonic surface, so we call this the curvature-induced electrostatic dipoles (CEDs). As one might have expected, large corrugations give larger CEDs, as shown in Figure 2d.

To further see the properties of CEDs, we plot the distributions of the electrostatic potential $-e\phi_0$ near the corrugated surfaces in the unit of electron-volt, as shown by the grey scale in Figure 2c. We also plot the iso-potential curves by rainbow colors for different values of the electrostatic potential (the potential contours are labeled in Figure 2c). The directions of the electric field generated from this potential are plotted in Figure 2c by the black arrows. We see that the CED generates an electric field near the corrugated surfaces, which can also be called as the fringe fields. This potential distribution further confirms the existence of CEDs for the corrugated surfaces. It should be mentioned that the considered corrugated surfaces above are globally charge neutral, namely $\int \sigma(x)\,\mathrm{d}x = 0$.

**Forces calculated using SC-HDM.** If we put two neutral nano-corrugated surfaces close to each other, the CEDs of different surfaces could interact via an electrostatic force between these two surfaces. To demonstrate this, we use two identical slabs separated with $\Delta d$, as shown in the right panel of Figure 1a. The parameters of the single slab are $r_a = 20\,\text{Å}$, $d = 20\,\text{Å}$, $A_u = 12\,\text{Å}$, and $A_b = 12\,\text{Å}$. Similar to previous calculations, we could use SC-HDM to calculate the ground states of electrons. To obtain the electrostatic force between these two slabs, we do the surface integral of the electrostatic stress tensor[37] for a single slab, and integrating surface is chosen to enclose the slab. In Figure 3a, we plot the electrostatic force acting on the bottom slab $\mathrm{Fy}_E$ as a function of $\Delta d$ by the blue line. As a consistency check,[38] we also calculate $\mathrm{Fy}_E$ by doing a volume integral of the Lorentz force density in the slab domain, as shown by the red dots in Figure 3a. The electrostatic stress tensor and the Lorentz force results are the same.

We note that that the forces in Figure 3a are positive, indicating that the electrostatic forces between these two slabs are attractive independent of the size of the gap. We plot the distribution of the electrostatic potential $-e\phi_0$ near the vacuum gap sandwiched between these two slabs in the unit of electron-volt, as shown by the grey scale in Figure 3c. The iso-potential curves and directions of electric fields are plotted by rainbow lines and black arrows, respectively, in Figure 3c. We see that the field lines go from the convex region of the bottom slab (positive charges) to the concave region of the upper slab (negative charges). This induces the attractive forces between these two slabs. The force is still nonzero when $\Delta d = 20\,\text{Å}$, indicating that the force is quite long-ranged.

The electrostatic forces change abruptly when $\Delta d < 9\,\text{Å}$. To understand the reasons, we calculate the quantum forces within the SC-HDM framework between these two slabs, originating from the internal energy of the electron gas. Since the electron energy functional for an electron density distribution *[n]* is frequently denoted as *G[n]* in the literature, we will call the corresponding force term $-n\nabla\left(\dfrac{\delta G}{\delta n}\right)$ the Fy$_G$ (see Methods for details). As it originates from the internal energy of the electron gas, Fy$_G$ will be zero when the electrons of these two slabs have no overlap so that they do not interact directly. In Figure 3b, we plot the Fy$_G$ force of the bottom slab by the green line. It can be seen that Fy$_G$ is essentially zero when $\Delta d > 10\,\text{Å}$, indicating that the electrons in the two slabs have no overlap. And hence in that region, the force is purely classical Coulombic in character. We note that we do not consider quantum fluctuation forces here. However, Fy$_G$ increases dramatically when $\Delta d < 10\,\text{Å}$, indicating that the electrons of these two slabs begin to interact with each other. Within this region, the two corrugated surfaces can be viewed as effectively connected by the tails of spilled-out electrons, so the forces between them are dominated by quantum internal forces. As long as there is no overlap of electron densities ($\Delta d > 10\,\text{Å}$), the force is Coulombic although the origin comes from the kinetic energy term of the density functional. We note that the magnitude of $Fy_E$ is about several hundred kilo-Pascal, which are comparable to atmospheric pressures. We also use our model to calculate the optical binding forces between these two slabs under plane wave illuminations.[34] Due to the field enhancement effect, light induced plasmonic forces acting on nanostructures are strong and can easily be two order of magnitude higher than ordinary radiation pressure (see section S1 in the Supporting Information for details) and are perceived to be useful for plasmon-based optical manipulations.[9-13,39] It is useful to compare the light-induced plasmonic force due to an external harmonic field with the corrugation-induced electrostatic force we are considering here. To this end, we compute the light induced force between the slabs (See Figure S1 in the Supporting Information). We found that for a laser with a power of $1.0 \text{mW}/\mu m^2$, the maximum attractive forces between the slabs are about 350 Pa at $\omega = 3.18 \text{eV}$ (tuned to the plasmonic resonance) under plane wave illumination in the normal direction, which is much smaller than the electrostatic forces $Fy_E$ predicted in Figure 3a (see Supporting Information for details). To reach the same magnitude as the corrugation-induced force, we need a laser intensity of $10^2 \text{mW}/\mu m^2$. We need a light power of $\sim 10^4 \text{mW}/\mu m^2$ to make the radiation pressure of an absorbing flat surface to produce the same force as the corrugation-induced force (see section S1 in the Supporting Information for details). This shows that the curvature-induced forces are not small comparing with typical optical forces used in optical manipulations.

We have demonstrated that curvature could induce attractive electrostatic forces between two neutral plasmonic objects because the positive and negative charge regions happen to be aligned with each other for the corrugation shown in Figure 1a. But there are other configurations in which the corrugation-induced forces could be repulsive. To see this, we consider the step-like corrugated surface configuration as shown in Figure 1b. The parameters used here are $r_1 = 10 \text{Å}$, $r_2 = 15 \text{Å}$, $d = 15 \text{Å}$, and $h_1 = 10 \text{Å}$. Then SC-HDM is employed to calculate the ground state of the system, and we plot the electrostatic force $Fy_E$ of the bottom slab as a function of $\Delta d$ in Figure 4b. The blue line and red dots are obtained by electrostatic stress tensor and Lorentz force density, respectively. They agree with each other. The most interesting feature in Figure 4b is that $Fy_E$ could change from an initially repulsive force to an attractive force when decreasing the interslab distance $\Delta d$. When $\Delta d > 11 \text{Å}$, the electrostatic forces are repulsive. In this case, the inter-slab forces are dominant by the convex regions of these two surfaces, which bear the same charge. To see this, we plot the electrostatic potential and electric fields in Figure 4c for $\Delta d = 15 \text{Å}$. We see that the field lines tend to be avoid-crossing inside the vacuum gap, indicating the electrostatic forces are repulsive. The electrostatic forces becomes attractive when $\Delta d < 11 \text{Å}$, after the protrusions are well inserted into the grooves. In this case, the convex protrusions are close to the concave regions, and opposite charges attract. We plot the electrostatic potential and electric fields in Figure 4a for $\Delta d = 9 \text{Å}$. In contrast with Figure 4c, it is clear that the field lines connect the convex region of one surface to the concave region of another surface, indicating the attractive electrostatic forces. It should be mentioned that the transition from repulsive forces to attractive forces occurs when $\Delta d \cong h_1$. This results from the competition between the convex-convex (concave-concave) interactions with the convex-concave interactions. When the convex-convex interactions are dominant, the electrostatic forces are repulsive, while convex-concave interactions favor attractive forces.

**Forces calculated for Na using LDA.** To further verify the existence of the curvature induced forces, we performed first-principles local density-functional calculations for sodium tip and step structures for comparison with SC-HDM results. The Na tip structure (triangular protrusions) and step structure (rectangular protrusions) are constructed using an ideal Na body-center cubic lattice as starting positions and the atomic coordinates are then relaxed using the Hellmann–Feynman forces. The final atomic configurations are shown in Figures 5a and 5b for an inter-slab distance of about $20 Å$, with the unit cell marked by black dotted lines. Periodic boundary conditions were applied in all three directions. The electrostatic forces per unit area as a function of inter-slab distance were calculated using the surface integral of the electrostatic stress tensor for Na-tip structure and Na-step structure are shown in Figures 5c and 5d, respectively. The electric field in the vacuum region is determined by the gradient of the Coulomb potential, which in turn is determined by the self-consistent charge density. The surface integral method can be justifiably applied only if the charge densities at the integral surface are negligible. In Figures 5c and 5d, the red solid dots mark the force calculated with the integration boundary set at the middle of the vacuum gap region ($y = Y_c$, as marked by the red lines inside the unit cell in panels a and b), while the blue crosses and green triangles show results calculated with the integration surface positioned at $y = Y_c \pm 1 Å$. If the electron density is negligibly small at the integration boundary, these three boundaries should give the same results. We see that the forces calculated with the three different boundaries do overlap for vacuum gap distances bigger than $13 Å$ for both structures so that the results are trustworthy as long as the vacuum gap between the slabs is bigger than $13 Å$. In Figures 5c and 5d, a force in the positive y-direction means an attractive force between the slabs. Figure 5c shows that the electrostatic forces between adjacent slabs are always attractive for the triangular tip corrugation while Figure 5d shows that there is initially a repulsion when the slabs are far apart for the square step corrugation, but the electrostatic interaction becomes attractive when the protrusions become inserted into the grooves. The results of first-principles calculation hence qualitatively agree well with that of SC-HDM. In particular, we see that for the case of Na-tip structure, the attractive forces decreases as the inter-slab distance increases, and the force is quite long-ranged and it is still nonzero when the inter-slab distance is more than $20\,\text{Å}$. For Na-step structure, the repulsive forces at large distances are observed in both the first-principles and the SCHDM calculations. In order to understand the mechanism producing the repulsive forces, we plot the electrostatic potential for the Na-step structures with $\Delta d = 12.1\,\text{Å}$ and $\Delta d = 31.1\,\text{Å}$ in Figures 6a and 6b, respectively. We see that in Figure 6a the potential increases monotonically from the convex region of the bottom step to the concave region of the upper step. This produces the attractive forces between the concave region and the convex protrusion. On the contrary, when the slabs are far apart, the corner of the convex protrusions are close to each other and electrostatic potential is of the same value in the convex regions of these two steps, as shown in Figure 6b, resulting in a repulsion.

**Forces calculated for W using LDA.** We showed that first-principles calculations and SC-HDM found the same qualitative results and the same mechanism for the curvature-induced attractions/repulsions for nano-corrugated Na surfaces. In this section, we consider tungsten (W), which is a refractory metal. The W step structure are constructed using an ideal W body-center cubic lattice as starting positions and the atomic coordinates are then relaxed using the Hellmann–Feynman forces. The final atomic configurations are shown in Figure 7a for an inter-slab distance of about $11\,\text{Å}$ and the electrostatic forces per unit area as a function of inter-slab distance $\Delta d$ were shown in Figure 7b. We see that the forces calculated with the three different boundaries do overlap for vacuum gap distances bigger than $11\,\text{Å}$ so that the results are trustworthy as long as the vacuum gap between the slabs is bigger than $11\,\text{Å}$. Figure 7b shows that the forces are repulsive initially when the slabs are far apart for the square step corrugation, but the electrostatic interaction becomes attractive when the protrusions intrude into the grooves.

The behaviors of the electrostatic force between nano-corrugated W surfaces are similar to that of nano-corrugated Na surfaces, and the underlying mechanism is also the same (see section S2 in the Supporting Information for details).

**Summary and Conclusions**

We showed that geometric curvature can induce local charges on corrugated plasmonic surfaces, with a convex curvature favoring a positive charge and concave curvature favoring a negative charge. On corrugated surfaces, these local charges form electrostatic dipoles with directions and magnitudes that depend on the geometric details and they generate electric fields near the surfaces. These corrugation induced dipoles are different from the usual interfacial electrostatic dipoles. The interactions between these electric fields from different corrugated surfaces induce an electrostatic force between neutral plasmonic surfaces. We show that this corrugation-induced force can be attractive or repulsive, depending on the details of the corrugation. The magnitude of this force is not small compared with other forces that can manipulate nano-objects, such as optical forces induced by reasonably strong external light sources. The underlying reason is the non-coincident distribution of nuclei charges and electron density, with the latter being smoother due to the need to minimize the kinetic energy. This effect is expected to exist in different metallic systems, and indeed we found it in prototypical simple metal (sodium, as described by both SC-HDM and LDA) and tungsten.

**Methods**

**Electrostatic stress tensor.** The electrostatic force $\mathbf{F}_E$ between the two corrugated surfaces in both the SC-HDM and first principles (VASP) approaches is obtained by doing the surface integral of the electrostatic stress tensor $\mathbf{T}$ on the surface $\partial\Omega$ enclosing the objects

$$\mathbf{F}_{\alpha,E} = \int_{\partial\Omega} \sum_{\beta=x,y,z} \mathbf{T}_{\alpha\beta} n_\beta \, dS,$$

where $\alpha, \beta$ denotes the components of certain vectors and tensors, $\mathbf{n}$ is the outward normal to the closed surface $\partial\Omega$, $dS$ is the infinitesimal area of the closed surface $\partial\Omega$, the electrostatic stress tensor is[37]

$$\mathbf{T}_{\alpha\beta} = \varepsilon_0 \left[ \mathbf{E}_\alpha \mathbf{E}_\beta - \frac{1}{2}(\mathbf{E}\cdot\mathbf{E})\delta_{\alpha\beta} \right],$$

and $\mathbf{E} = -\nabla\phi_0$ is the electrostatic fields generated by all the charges. $\phi_0$ is the overall Coulomb potential by all the electrons and nuclei, and it can be calculated within both SC-HDM and VASP.

The integration surface $\partial\Omega$ should be chosen to lie within the charge free region (as shown in Figures 3 and 5) in order to obtain the total electrostatic force, otherwise, the integral will include the internal forces of the systems.

**Self-consistent hydrodynamics model.** Within the SC-HDM, the electron gas in the ground state can be fully described by the electron density $n_0(\mathbf{r})$, and electrostatic potential $\phi_0(\mathbf{r})$. The equations that need to be solved numerically are,[34]

$$-n_0 \nabla \left( \frac{\delta G}{\delta n} \right)_0 + q_e n_0 \mathbf{E}_0 = 0,$$

$$\nabla^2 \phi_0 = \frac{q_e}{\varepsilon_0}(n_j - n_0), \tag{1}$$

where $q_e = -e$ is the charge of electron $(e > 0)$, the static electric field $\mathbf{E}_0$ is $\mathbf{E}_0 = -\nabla \phi_0$, and the energy functional $G[n(\mathbf{r})]$ is choose as[34]

$$\frac{\delta G}{\delta n} = \frac{\hbar^2}{2m_e}(3\pi^2 n)^{2/3} + \frac{\lambda_\omega \hbar^2}{4m_e}\left[\frac{\nabla n \nabla n}{2n^2} - \frac{\nabla^2 n}{n}\right]$$
$$-0.0588 \times \frac{4e^2}{3\varepsilon_0}n^{1/3} - \frac{4e^2}{3\varepsilon_0}\frac{0.035}{0.6024 + 7.8a_H n^{1/3}}n^{1/3},$$
$$+ \frac{7.8a_H n^{2/3}}{3}\frac{0.035}{(0.6024 + 7.8a_H n^{1/3})^2}\frac{e^2}{\varepsilon_0}$$

In this work, we choose $\lambda_\omega = 0.12$.[34] The positive charge background $n_j$ is taken to be the standard jellium model, namely $n_j = n_{ion}$ inside the metal, and $n_j = 0$ outside the metal. Substituting $\mathbf{E}_0 = -\nabla \phi_0$ into the above equations could obtain

$$\left(\frac{\delta G}{\delta n}\right)_0 + q_e \phi_0 = \mu, \qquad (2)$$

where $\mu$ is the chemical potential of the electron system. We need to specify the total number of electrons. The particle is charge neutral, implying the following constraint

$$\int_\Omega d\mathbf{r}\, e[n_+ - n_0] = 0. \qquad (3)$$

Equations (1)-(3) collectively determine the ground state charge densities of the corrugated surfaces.[34] As the basic variables of the SC-HDM model are the electron densities instead of electronic wave functions, the method is computationally more efficient than fully ab-initio methods such as TD-DFT. This method can handle the electron spill-out effect near metal surfaces which is essential for the present problem.

**Density functional calculations.** The accurate frozen-core full-potential projector augmented-wave (PAW) method was used,[40] as implemented in Vienna ab-initio simulation package (VASP).[41-43] For the exchange and correlation energy, we adopted the Perdew-Burke-Ernzerhof generalized gradient approximation.[44] The experimental lattice constant ($a = 4.23 \text{Å}$ for Na and $a = 3.19 \text{Å}$ for W) is used as the starting position in this calculation. All atoms were fully relaxed until the maximum magnitude of the Hellmann–Feynman force acting on the atoms was smaller than $0.02 \text{ eV/Å}$. The k-points sampling was set at 2 x 18 x 2.

**ASSOCIATED CONTENT**

**Supporting Information**. The Supporting Information is available free of charge on the website at DOI:

Absorption spectrum and optical binding forces of corrugated sodium slabs (S1).

Electrostatic potential distributions of tungsten step structures (S2).

**AUTHOR INFORMATION**

**Corresponding Author**

* E-mail: phchan@ust.hk.

**Author Contributions**

C.T.C. conceived the project. K.D. performed the SC-HDM calculations. H.H. and T.C.L. performed the first principle calculations. All the authors contribute to the writing of the manuscript, and have given approval to the final version of the manuscript.

**Notes**

The authors declare no competing financial interest.

**ACKNOWLEDGMENTS**

This work is supported by Research Grants Council, University Grants Committee, Hong Kong (Grants No. AOE/P-02/12). H. Hu and T. C. Leung acknowledge NCTS, financial support from NSC of Taiwan under Grant No. NSC-105-2112M-194-006, and the computer time at the National Center for High-performance Computing.

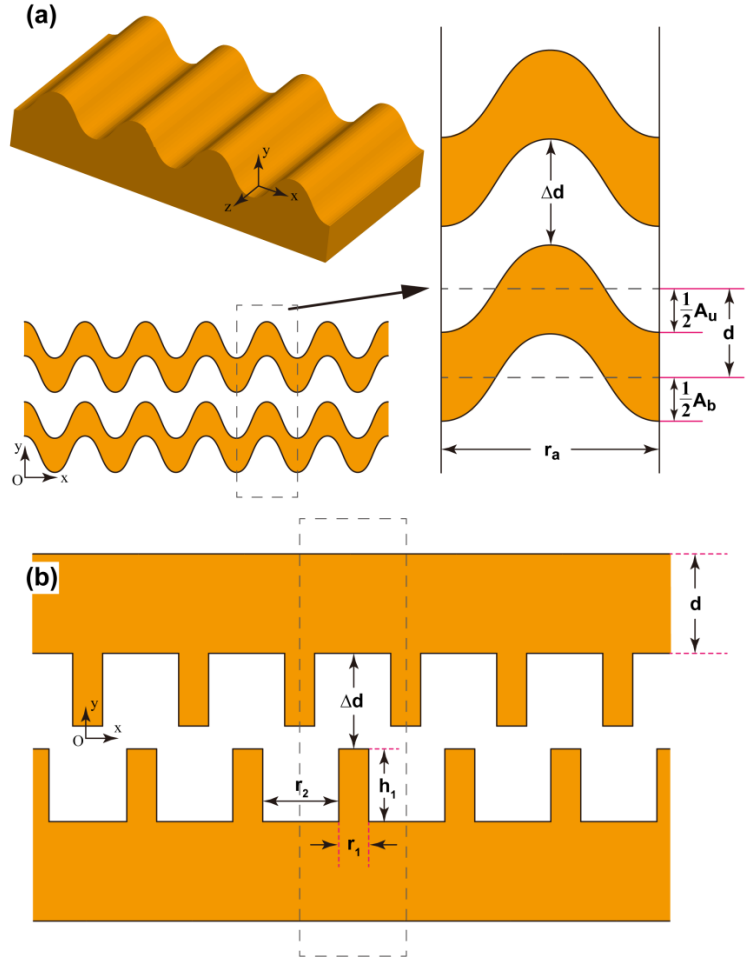

**Figure 1.** (a) Schematic picture of a corrugated surface (upper left panel). Cross section view of the plasmonic slabs under consideration (lower left panel) and the calculation domain for a single unit cell with lattice constant $r_a$ (right panel). The thickness of each slab is $d$, the distance between them is $\Delta d$ and the corrugation is defined by a the function $0.5 A_{u,b} \cos(\pi x / r_a)$. (b) Schematic picture of two corrugated surfaces with rectangular protrusions, with various parameters defined in the figure.

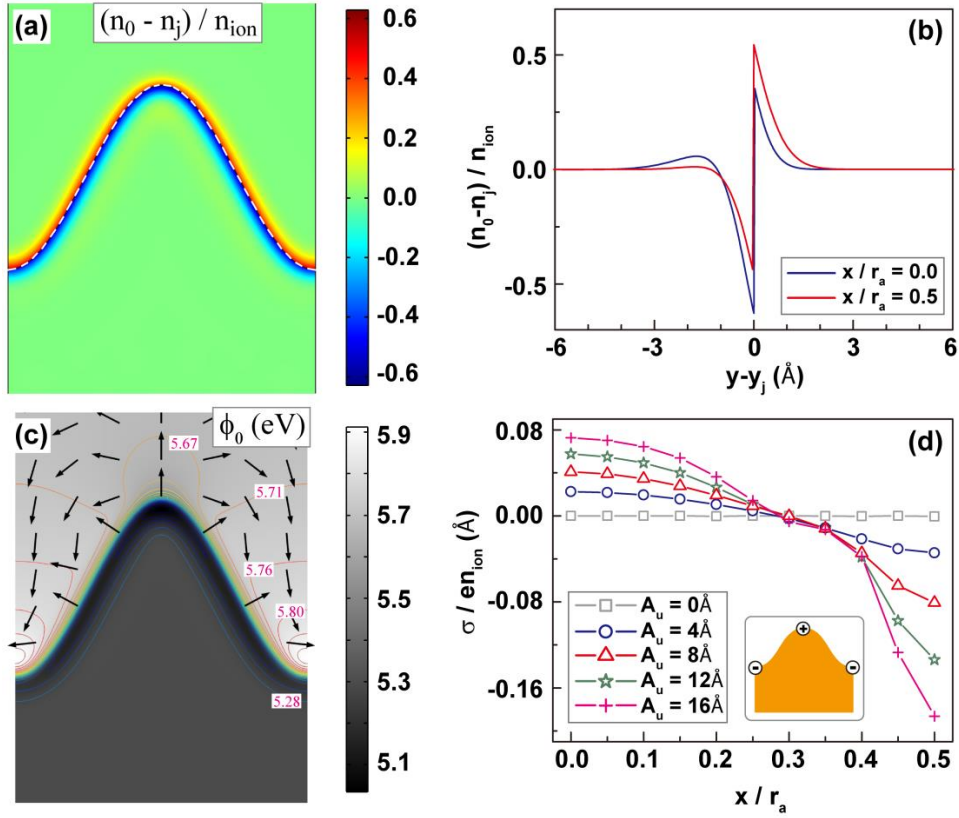

**Figure 2.** (a) Normalized charge distributions $(n_0 - n_j)/n_{ion}$ of a cosine-function corrugated surface with $r_a = 20\,\text{Å}$, $d = 50\,\text{Å}$, $A_u = 12\,\text{Å}$, and $A_b = 0\,\text{Å}$. The white dotted line marks the boundary of the jellium background. (b) The normalized charge distributions in the convex ($x/r_a = 0.0$) and concave ($x/r_a = 0.5$) directions are plotted by blue and red lines, respectively. The zero of the horizontal axis marks the jellium boundary. (c) The electrostatic potential for electrons is plotted by grey scale in the unit of eV. The equal potential surfaces are plotted by rainbow colors. The black arrows represent the directions of the electric field. (d) Surface charge densities as a function of position $x/r_a$ for different amount of corrugations $A_u$.

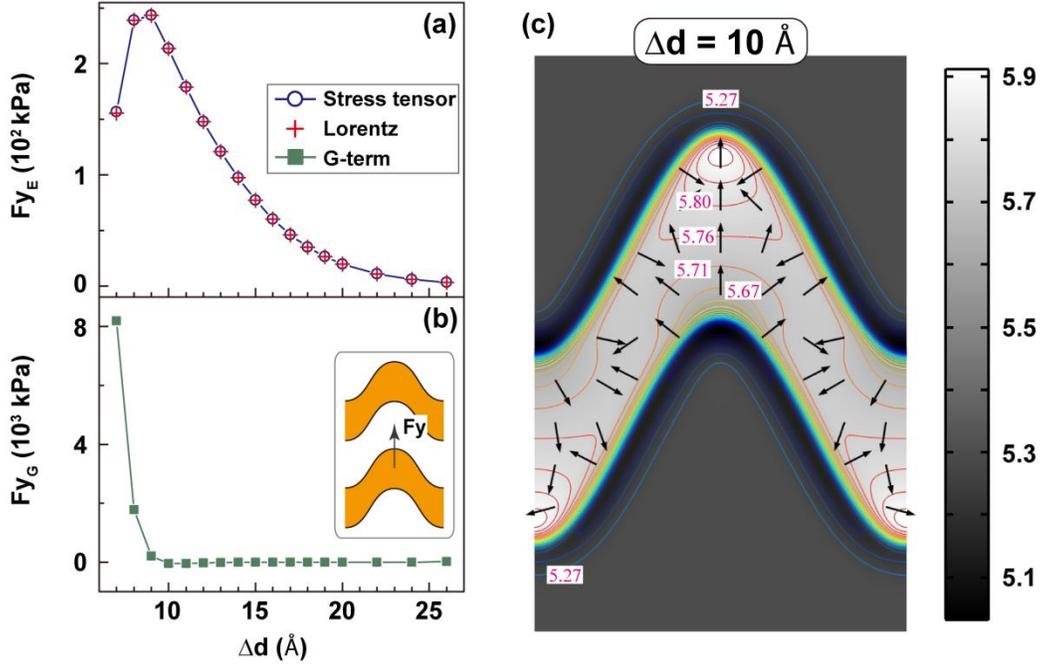

**Figure 3.** The electrostatic forces and the "quantum" forces acting on the bottom slab as a function of $\Delta d$ are plotted in (a) and (b), respectively. The other parameters used are $r_a = 20\,\text{Å}$, $d = 20\,\text{Å}$, $A_u = 12\,\text{Å}$, and $A_b = 12\,\text{Å}$. The blue lines and red dots in (a) are calculated by the surface integral of electrostatic stress tensor and volume integral of Lorentz force density, respectively. (c) The distributions of electrostatic potential for the electrons near the vacuum region are plotted in grey scale in the unit of electron-volt at the gap size of 1.0nm. The equal potential surfaces are plotted by rainbow colors with electrostatic potential marked on it. The black arrows represent the directions of the electric field.

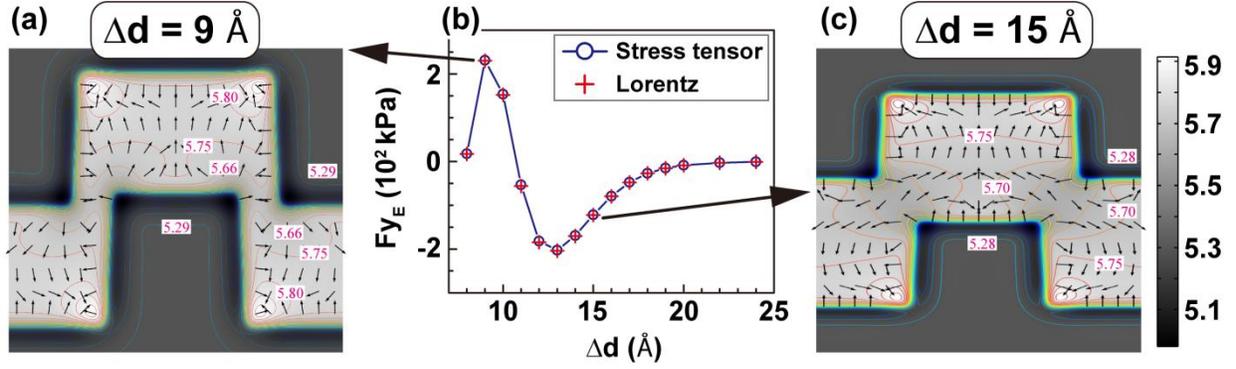

**Figure 4.** The electrostatic forces acting on the bottom slab with rectangular protrusions as a function of $\Delta d$ are plotted in the middle panel (b). The other parameters used are $r_1 = 10\,\text{Å}$, $r_2 = 15\,\text{Å}$, $d = 15\,\text{Å}$, and $h_1 = 10\,\text{Å}$. The blue lines and red dots are calculated by the surface integral of the electrostatic stress tensor and volume integral of Lorentz force density, respectively. The distributions of electrostatic potential for the electrons near the vacuum region are plotted by grey scale in the unit of eV for $\Delta d = 9\,\text{Å}$ and $\Delta d = 15\,\text{Å}$ in (a) and (c), respectively. The equal potential surfaces are plotted by rainbow colors, and the black arrows represent the directions of the electric field.

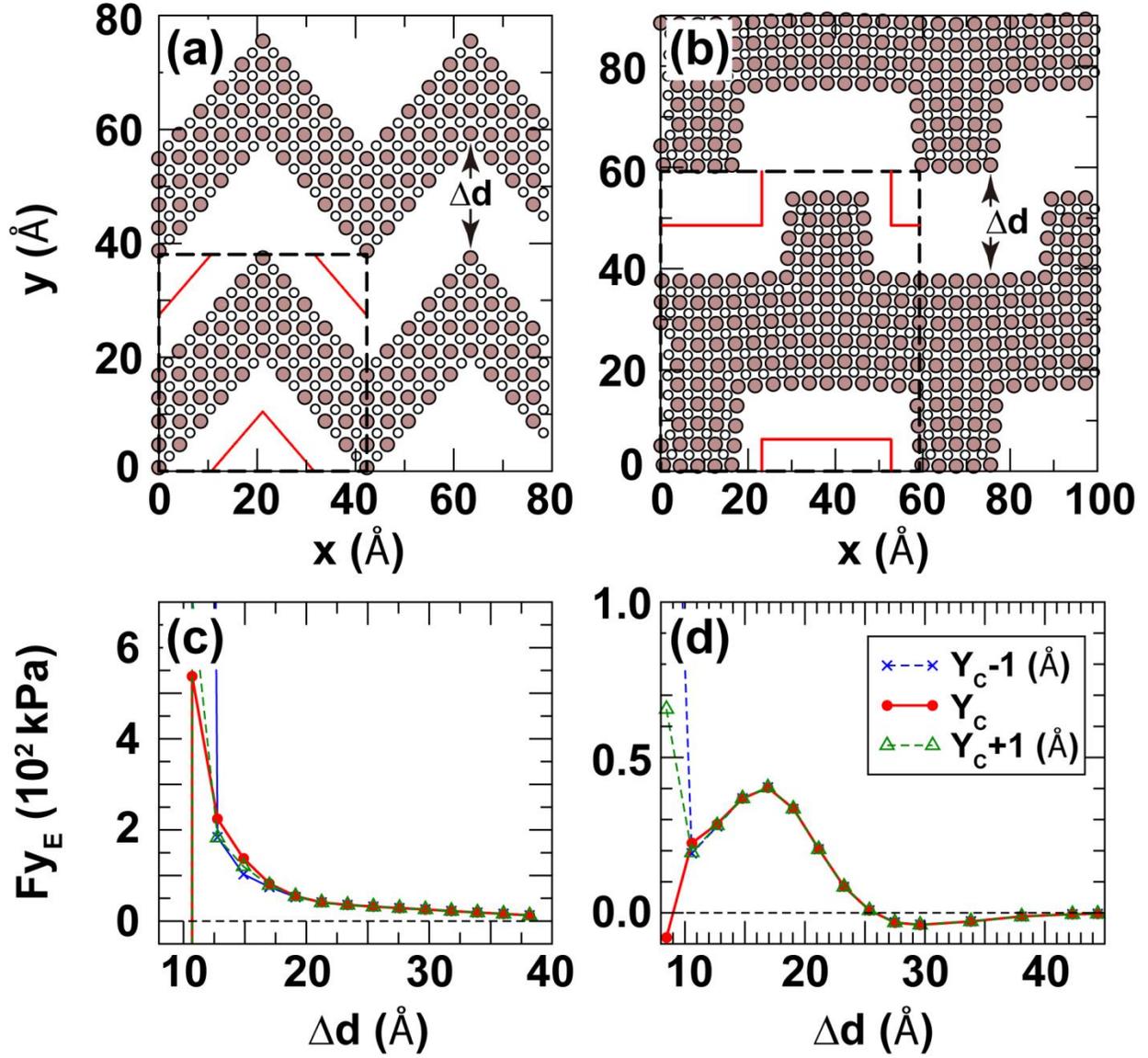

**Figure 5.** Picture of (a) Na-tip structure and (b) Na-step structure for first-principles calculations. Filled brown circles denote Na atoms in the $z = 0$ plane and open black small circles represent Na atoms in the $z = 0.5a$ plane. The black dashed lines mark the primitive cell boundaries in the x-y directions. The red lines mark the integral surface for electrostatic forces between the slabs (at the center of vacuum gap $y = Y_c$). The electrostatic forces per unit area calculated by the surface integral of the electrostatic stress tensor for Na-tip structure and Na-step structure are shown in (c) and (d), respectively. The solid red circles show results calculated at the center of the vacuum gap, the blue crosses and green open triangles denote the results calculated at the position 1.0 Å below and above the center of the vacuum gap, respectively.

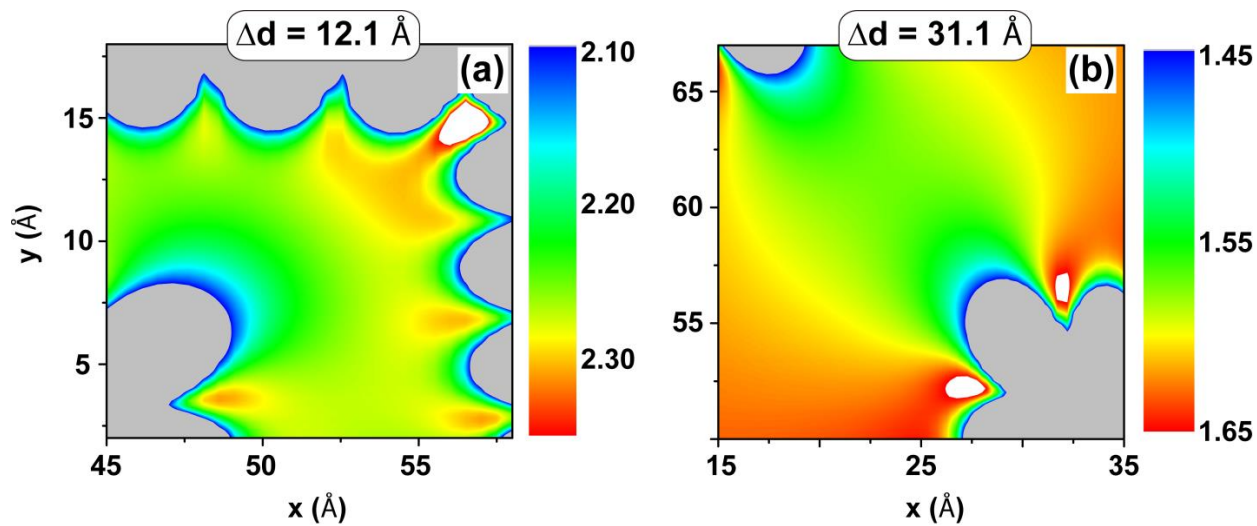

**Figure 6.** Distributions of electrostatic potential in the unit of eV for the Na-step structures with (a) $\Delta d = 12.1 \text{Å}$ and (b) $\Delta d = 31.1 \text{Å}$ calculated within first-principles calculations. All the other parameters are the same with Figure 5d. The grey areas correspond to domains where the potential is off-scale due to atomic charges.

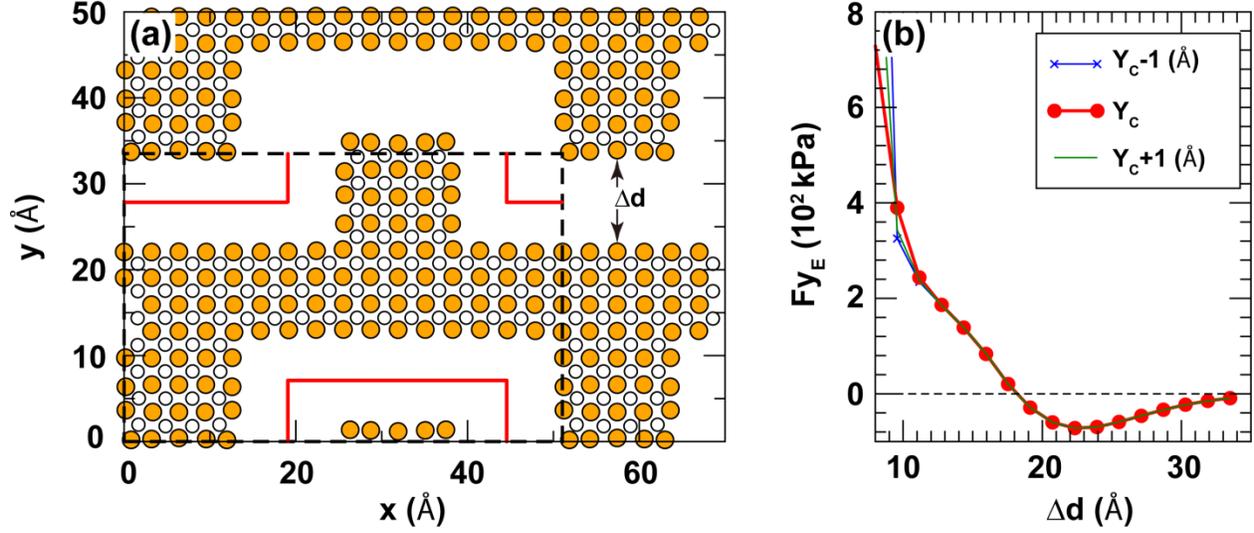

**Figure 7.** (a) Picture showing W-step structure. Filled orange circles denote W atoms in the $z = 0$ plane and open black small circles represent W atoms in the $z = 0.5a$ plane. The black dashed lines mark the primitive cell boundaries in the x-y directions. The red lines mark the integration surface for electrostatic forces between the slabs (at the center of vacuum gap $y = Y_c$). The electrostatic forces per unit area calculated by the surface integral of the electrostatic stress tensor for W-step structure are shown in (b). The solid red circles show results calculated at the center of the vacuum gap, the blue crosses and green open triangles denote the results calculated at the position 1.0 Å below and above the center of the vacuum gap, respectively.

# Supporting Information --- Nano-corrugation induced forces between electrically neutral plasmonic objects

*Kun Ding,[†] Han Hu,[‡] T. C. Leung,[‡] and C. T. Chan[\*,†]*

[†] Department of Physics and Institute for Advanced Study, The Hong Kong University of Science and Technology, Clear Water Bay, Hong Kong

[‡] Department of Physics, National Chung Cheng University, Chia-Yi 62101, Taiwan

\* Corresponding E-mail: phchan@ust.hk

**S1. Absorption spectrum and optical binding forces of corrugated slabs**

To compare the curvature-induced electrostatic force with the light induced forces on plasmonic surfaces, we employ SC-HDM to calculate the response of corrugated slabs shown in Figure 3 under external electromagnetic waves.[1] The slab parameters used are the same as Figure 3. The incident light is a plane wave with $\mathbf{k}//\hat{y}$ and $\mathbf{E}//\hat{x}$, as shown in the inset of Figure S1.

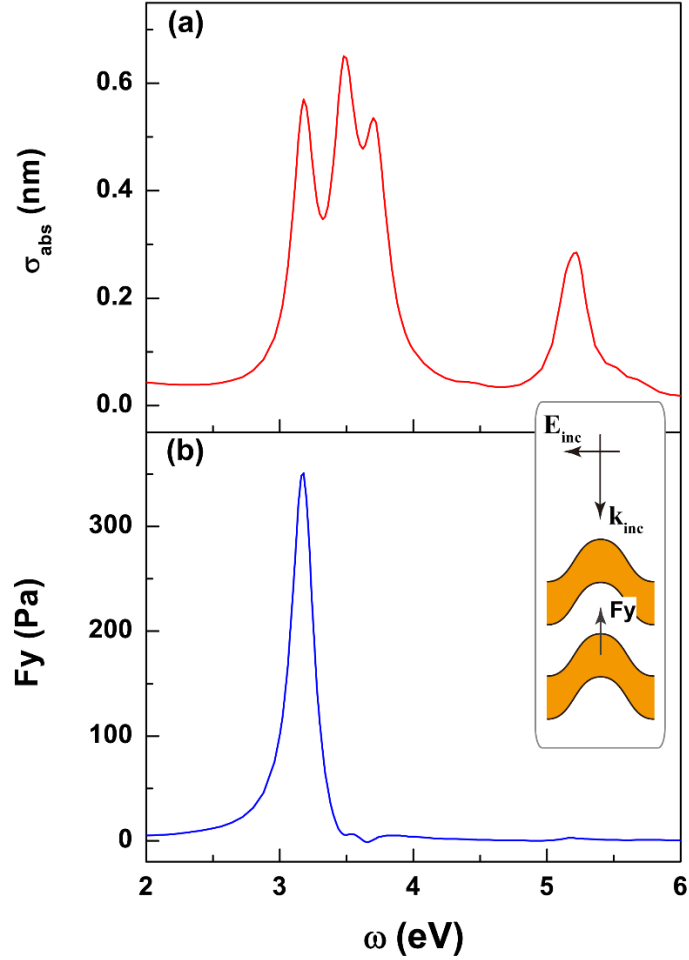

**Figure S1.** (a) Absorption spectrum and (b) optical binding force acting on the bottom slab of the corrugated surface system under plane wave illumination. The structure of corrugated slab, the incidence direction, the light polarization and the direction of the optical force are shown in the inset. The beam power is 1.0mW/μm$^2$.

The absorption spectrum in Figure S1(a), calculated using SC-HDM, shows that the corrugated slab system have several plasmonic resonances. We calculate the optical force induced by a plane wave, with k-vector and polarization as shown in the inset of Figure S1. The intensity of the continuous wave (CW) light source is 1.0mW/μm$^2$. The optical force, in the direction shown in the inset, is plotted in Figure S1(b). As the sign is positive, the force is attractive between the layers and in the same direction as the corrugation-induced force for this particular corrugation configuration. We see that the maximum light induced binding force in the spectrum appears at the resonance at $\omega = 3.18\text{eV}$, and the magnitude of the maximum optical pressure is about 350 Pa. So in order to make the optical binding force comparable with the corrugation-induced electrostatic force shown in the main text, the incident beam power needs to reach $10^2\,\text{mW/μm}^2$, which is fairly strong for a CW light source.

We can also make reference to the radiation pressure of a CW plane wave acting on a flat surface. Under plane wave incidence, the optical pressure for an object with a geometric cross-section $A$ is $\mathbf{F}/A = \dfrac{C_{ext}}{Ac}\langle\mathbf{S}\rangle$,[2,3] where $C_{ext}$ is the extinction cross-section and $\langle\mathbf{S}\rangle$ is the time averaged Poynting vector of the incident light. Hence for a perfectly absorbing flat surface ($C_{ext} = A$), the incident light power needs to reach $10^4\,\text{mW/μm}^2$ in order for the magnitude of optical pressure to $10^2\,\text{kPa}$. A perfectly reflecting metallic surface will have twice the radiation pressure. If we compare the light induced force acting on a corrugated plasmonic surface with that acting on an absorbing/reflecting flat surface, we see that the optical force on corrugated plasmonic surface is two orders of magnitude bigger. This is not surprising because of the field enhancement effect, and the corrugation-induced force discussed in this paper is roughly equivalent to the optical force induced on corrugated plasmonic surface by a fairly strong light source tuned to the resonance frequency.

**S2. Electrostatic potential distributions of tungsten step structures**

To show the mechanism of attractive and repulsive electrostatic forces, we plot the distributions of the electrostatic potential obtained in VASP for W step structures in Figure S2.

We show the attractive force case ($\Delta d = 11.165 \text{Å}$) and repulsive force case ($\Delta d = 22.33 \text{Å}$) in Figures S2(a) and S2(b), respectively. We see that the potential increases monotonically from the convex region of the bottom step to the concave region of the upper step, and hence produce the attractive forces. For the repulsive force case, the potential of the convex regions of these two steps is of the same value, which causes the repulsive forces. Figure S2 shows the mechanism of W step structures is the same as the sodium cases shown in Figures 3, 4, and 6 of the main text.

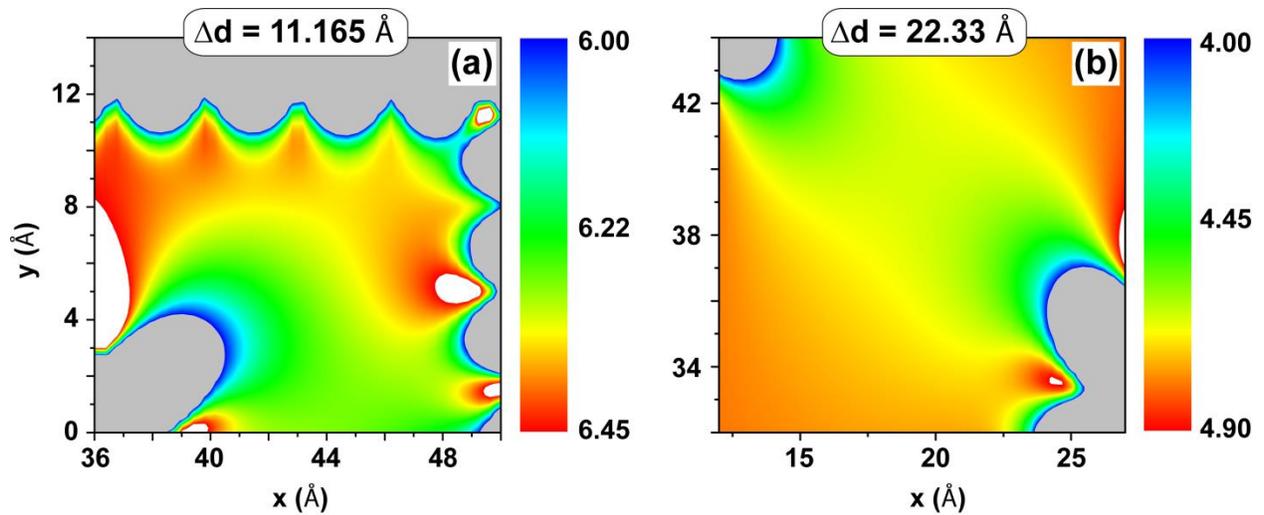

**Figure S2.** Distributions of electrostatic potential in the unit of eV for the W-step structures with (a) $\Delta d = 11.165 \text{Å}$ and (b) $\Delta d = 22.33 \text{Å}$ calculated within first-principles calculations. All the other parameters are the same with Figure 7 in the main text. The grey areas correspond to domains where the potential is off-scale due to atomic charges.